\documentclass[jphysg,12pt]{article}
\usepackage{graphicx}
\usepackage{epsfig}
\begin{document}
\baselineskip 20pt plus .1pt  minus .1pt
\pagestyle{plain}
\voffset -2.0cm
\hoffset -1.0cm
\setcounter{page}{01}
\rightline {}
\vskip 1.0cm

\begin{center}
{\Large {\bf Gamma Rays from the Galactic Centre}}
\end{center}
\begin{center}
A.D.Erlykin $^{1,2}$, A.W.Wolfendale $^2$
\end{center}

\begin{center}
\small (1) P. N. Lebedev Physical Institute, Moscow, Russia\\
(2) Department of Physics, University of Durham, Durham, UK
\end{center}

\begin{abstract}
Recent results from the HESS gamma ray telescope have shown the presence of
both a diffuse, extended, flux of gamma rays above $\sim$0.4 TeV and
discrete sources in and near the Galactic Centre. Here, we put
forward a possible explanation in terms of the diffusion of cosmic
ray protons from a succession of supernova remnants (~SNR~) in the 
SgrA* region of the Galaxy plus a contribution from SNR in the
rest of the Galactic Centre Region, to be called the Galactic Centre
Ridge (~GCR~). Protons are favoured over electrons because the $mG$
magnetic fields in the Region will attenuate energetic electrons severely. 

Prominent features are the need for 'anomalous diffusion' of the
protons in the whole region and the adoption of low efficiency for
SNR acceleration in the high density regions. The latter is related
by us to the well-known low 'cosmic ray gradient' in the Galaxy.

A corroborating feature is the close correlation of inferred cosmic ray
intensity with the smoothed intensity of 5 GHz radio radiation. We
attribute this to the presence of the SNR in the GCR. 
\end{abstract}

\section{The Galactic Centre}

It is well known that the Galactic Centre Region (~GCR~) has
remarkable properties, and it is not surprising that the Cosmic Ray 
(~CR~) view has singular interest.

Concerning molecular gas, some 10\% of such gas in the Galaxy is
to be found in the central 200 pc radius (~Morris and Serabyn, 1996,
to be denoted I~). The magnitude is quoted as $(5-10)\cdot
10^{7}~M_{\odot}$ (I), with probably  ~$4.4 \cdot 10^{7}M_{\odot}$ in the region
of concern here, $r \leq 200~pc$ (Tsuboi et al., 1999). The mean
density is very high, $\langle n \rangle > 10^{4}~ cm^{-3}$, in many of
the clouds and the filling
factor similarly high: $\sim 0.1$. The mean temperature is
elevated, being typically 70K (~in the range 30K - 200K~) (I).
Surrounding the molecular clouds is a very hot plasma of
`temperature' 10-15 keV (I), a value which is so high as to need
the presence in the past of very energetic activity, such as a
giant Galactic Centre explosion and/or the explosion of many SN within
the central parsec over the past $10^4 - 10^5~$years (I).

Many measurements have indicated the existence of a magnetic field
of several mG, with both tangled and smooth components. A prominent
feature is the presence of field 'tubes' nearly perpendicular to
the Galactic Plane (I).

SgrA* comprises a compact central object containing a radio source
with a flat spectrum. It is almost certain that a black hole of
mass $(2-3) \times 10^{6}~M_{\odot}$ is involved (I).

Of many other unusual features associated with the GCR, mention 
should be made of the high velocity winds (~500 -
1000 $kms^{-1}$~) (~Breitschwerdt et al.,2002~; V\"{o}lk and
Zirakashvili,2004~) associated 
particularly with the region SgrB2 (~eg Sunyaev et al., 1993~).

Particular remarks should be directed to the 'energetics' of the GCR;
we are mindful of the fact that, locally, the energy densities of CR,
starlight, magnetic fields and gas motion are all equal at $\simeq 0.5
eV cm^{-3}$ (~Wolfendale, 1983~). In the GCR we estimate, from the published 
data given in (I), that the energy densities of plasma, the far-infra-red radiation, 
magnetic fields and gas motion are all several thousand $eV cm^{-3}$ (~indeed, the
magnetic energy density will be higher still over limited volumes~).

The features listed above all lead to particular interest in the
gamma ray signal from the general region of the Galactic Centre.
Early work by SAS II (~Thompson et al.,1976~), COSB
(~Mayer-Hasselwandler et al.,1982~) and CGRO (~Mayer-Hasselwandler et
al.,1998; Hartmann et al.,1999~) in the 100 MeV region showed that
the CR intensity there was probably not very different from that
locally, but the angular resolution was too poor for
detailed study. More recently, CANGAROO (~Tsuchiya et al.,2004~), VERITAS
(~Kosack et al.,2004~), HESS (~Aharonian et al.,2004~) and MAGIC
(~Albert et al.,2006~) have observed the GC in sub-TeV and TeV gamma rays. Now,
HESS, with its superior angular resolution (~$\simeq 0.1^\circ$~) coupled
with its higher energy range (~ from hundreds of GeV to tens of TeV~),
offers the possibility of another window on this dramatic region
(~Aharonian et al., 2006~) and we shall base our analysis mainly on this
latest work. 

\section{The cosmic ray aspect}

Over the years we have examined the implications of cosmic ray
acceleration in supernova remnants (SNR) for a wide variety of
CR phenomena. A strong case has been made for such
acceleration providing the bulk of CR up to $10^{15}~eV$ (eg
Erlykin and Wolfendale, 2001) and perhaps - for a special class of
SNR - beyond (Biermann, 1993; Sveshnikova, 2003; Erlykin and
Wolfendale, 2005~).
 
Gamma ray astronomy provides a proxy indicator of CR in distant
parts of the Galaxy by way of gamma ray production in CR -
interstellar gas collisions and CR interaction with magnetic and photon
fields. A problem is that the nature of the
CR (nuclei or electrons) is not known, {\em a priori}, and also there are
often difficulties with the determination of the necessary column
density of gas. Nevertheless, in the absence of better ways of
studying the origin of CR, we proceed, noting particularly that high
energy electrons - those necessary for TeV gamma rays - will be
severely inhibited by the $mG$ magnetic fields, whether they be primary
or secondary (~in any event, the flux of secondary electrons well away
from the source will be very small~).

HESS has given contours of the TeV gamma ray emission from the
exciting Galactic Centre Region (~GCR~), specifically, that bounded
by $-2^{o} <l< + 2^{o} , -1.2^{o} <b< 1^{o}$
(~http://www.mpi-hd.mpg.de/hfm/HESS~). 
The `map' is
characterized by `point source' emission from the `compound SNR'
$G 0.9 + 0.1$ and from SgrA*, very close to the nominal GC.

\section{The basic data}

\subsection{The gamma ray map}

Following the analysis of the HESS group, we consider the region:
$|l|<1.6^{o}$ and $|b|<0.3^{o}$.
Figure 1 shows the profile of the `intensity' (actually the
counts) as a function of longitude (~Aharonian et al.,2006~). The result of subtracting
the two point sources ($G~0.9 + 0.1$ and Sgr A*) is indicated. The
workers used the known point spread function (PSF) for this
subtraction. The background has been also subtracted.

The authors point out that the diffuse gamma-ray emission diminishes with longitude
in the studied region and fades away at $|l| > 1.2^\circ$ inspite of the fact that 
there is a 
substantial amount of gas at least at positive longitudes of $l > 1.2^\circ$. They 
explain this profile assuming that gamma rays are produced in CR - gas collisions
and CR have a gaussian distribution around the GC with a best fit width of 
$0.8^\circ$. 

In what follows we draw the physical scenario of the observed features.  
We examine the possibility that our SNR model has validity here, too, viz that SNR are 
the sources and that the CR propagate in a diffusive manner. Specifically, we assume 
that many SNR occurred near the
GC during the last $10^4 - 10^5$ years, (~and extending back, perhaps,
even longer~) as already suggested. As in the HESS group scenario the gamma rays came
 from the interactions of CR protons from SNR with the ambient high density gas, 
essentially all the gamma rays from SgrA* coming from the burst of SNR, the remainder
(~from the GCR~) coming either from the burst of SNR or the 'conventional SN' in the 
GCR (~which would surely be expected in view of the considerable amount of stellar 
activity~).
\begin{figure}[htp]
\begin{center}
\includegraphics[height=7cm,width=12cm]{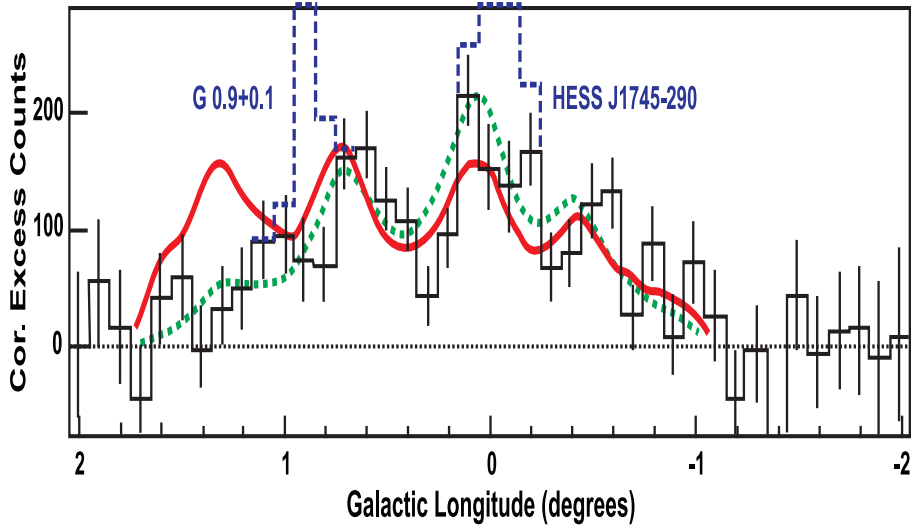}
\caption{\footnotesize The HESS results on gamma rays from the
Galactic Centre Region (~Aharonian et al., 2006~). The histogram gives the 
profile of gamma ray intensity above 0.38 TeV for $|b| <
0.2^\circ$. Two 'point' sources have been removed, as indicated by
dashed lines. The solid line is the column density of molecular
hydrogen between the same latitude limits normalised to the same total
area. The dashed line shows the gamma-ray flux expected if the CR
density distribution can be described by a Gaussian centred at $l =
0^\circ$ and with rms 0.8$^\circ$ (~following the analysis of the HESS
group~).}
\label{fig:fig1}
\end{center}
\end{figure}

\subsection{The distribution of target gas and the inferred CR
intensity}

Many measurements have been made of the distribution of gas in the
GC region, most notably using CO (eg Bania et al., 1977, Oka et
al., 1998) and CS ( Tsuboi et al., 1999). An apparent feature,
common to all analyses, is the very high density of gas ($H_{2}$) in
the region. As the last authors have remarked, there
are peculiar structures here which are `presumably related to the
unique activity in the GC region'. The HESS group used the CS data
to give the consequent column density of molecular hydrogen and we
have done the same, although we have applied a correction for the loss
of lower density molecular gas from the work of Dame et al.,2001 for 
the region $|l| < -1^\circ$.

At the beginning we tried to find the general characteristics of CR in the GC region.
Unlike the HESS group's assumptions about the gaussian distribution of CR around the GC
 we adopted a uniform distribution in this area and started with the local value of 
the CR intensity. Using our emissivity of the yield of gamma rays per hydrogen atom for
 the local CR spectrum (Erlykin and Wolfendale, 2003) we have derived $R(l)$ - the
ratio of the observed/expected gamma ray intensity with the result shown in Figure 2. 
If the observed gamma rays originate indeed from CR - gas collisions the gas 
distribution is eliminated from $R(l)$ and this ratio gives the actual CR longitudinal 
profile compared with the assumed uniform distribution. The possible contribution of 
electrons is ignored in this assumption following the arguments in \S2.

The mean value of $\langle R(l) \rangle$ averaged over the interval of 
$-1.4^\circ < l < 1.4^\circ$ is $2.44\pm 0.41$ for the total flux including the central
 source. Without the central source it is $\langle R(l) \rangle = 1.83\pm 0.15$.
It is with the interpretation of this plot that
we shall be mainly concerned.
\begin{figure}[htb]
\begin{center}
\includegraphics[height=15cm,width=8cm,angle=-90]{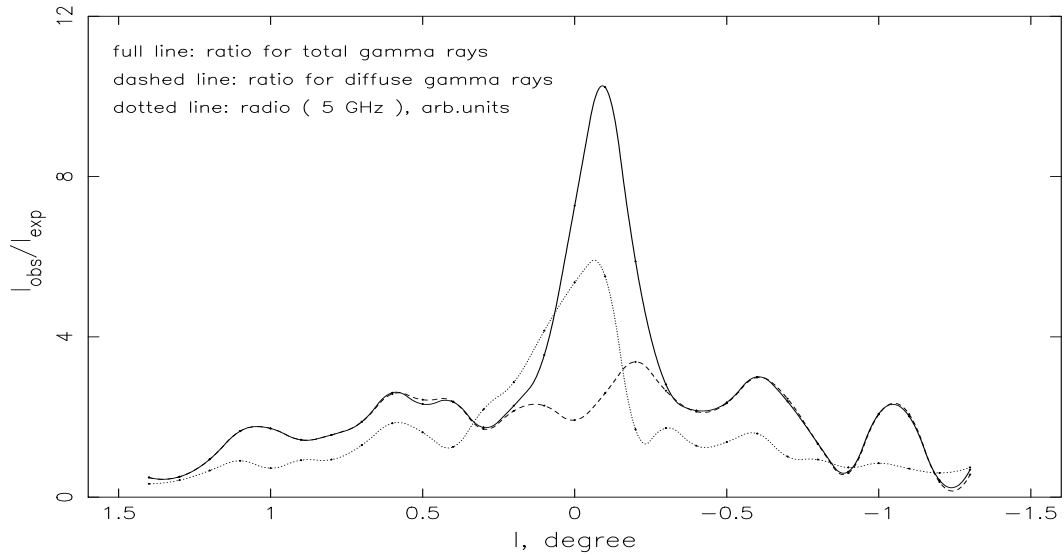}
\caption{\footnotesize Ratio of observed to expected gamma ray
intensity for $E_\gamma > 0.38$ TeV, the expectation being that the CR
intensity is the same as that locally; the ratio is thus that of the
proton intensity to that locally: $R(l)$. The full line shows the ratio
for the total gamma-ray intensity, the dashed line - the same, but with
the central source subtracted, the dotted line - the profile of the radio
intensity in arbitrary units. The contribution from CR electrons is
ignored following the arguments in \S2. Inspection of the basic data in
Figure 1 shows that most of the peaks are significant at the 2-3 standard 
deviation level.}
\label{fig:fig2}
\end{center}
\end{figure}

\subsection{The radio map}

Of likely relevance is the radio map for the same region.
Figure 2 shows the average intensity over $|b|<0.3^{o}$ derived by us
from the 5 GHz data of Altenhof et al., 1979, that, for 10 GHz is very
similar (~Handa et al.,1987~). Its relevance will be
considered briefly here and in more detail in \S5.3. 

Inspection of Figure 2 shows that there is a correlation between the
excess CR intensity and the radio emission intensity and this is quantified in
Figure 3. The slope of the straight line fit is 0.96$\pm$0.11 and the
corrrelation coefficient is 0.85; there is thus strong evidence for
linearity.
\begin{figure}[htb]
\begin{center}
\includegraphics[height=15cm,width=8cm,angle=-90]{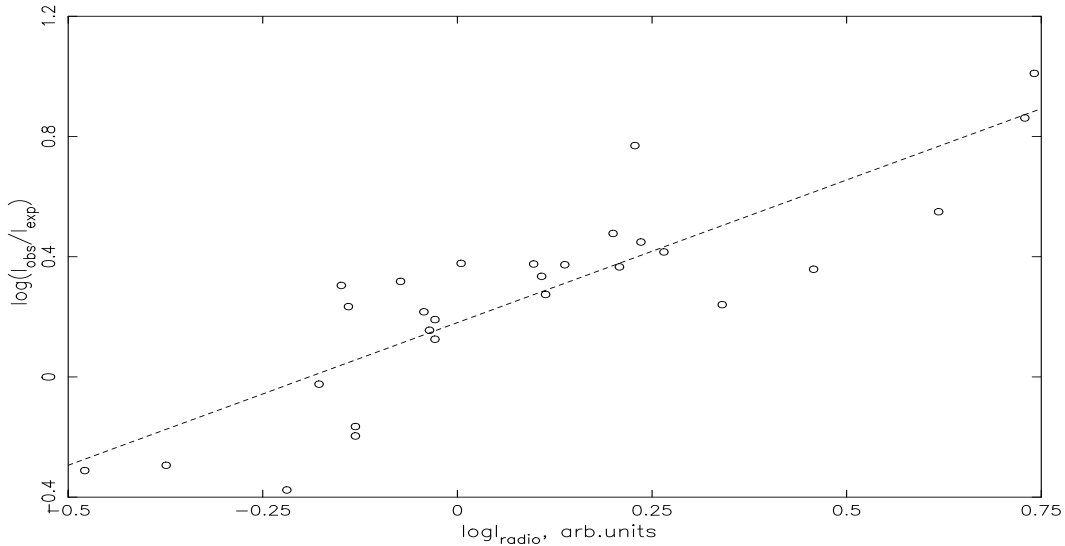}
\caption{\footnotesize Correlation of the CR intensity (~as distinct
from the gamma-ray intensity~) with the
integrated radio intensity (~at 5 GHz~) for $|b| < 3^\circ$. The
results relate to the GCR.~}
\label{fig:fig3}
\end{center}
\end{figure}

Although some of the radio photons come from discrete sources
(~'young' pulsars, etc~) and SNR at the `centers' of the condensed contours many
come from surrounding regions and are due to CR electrons (~from old
pulsars, SNR etc~) 
undergoing synchrotron radiation in the strong magnetic fields in
the region, together with a 'thermal contribution', which in fact is
greater than the non-thermal (~ie from synchrotron radiation~) at this
frequency.

In fact, the thermal contribution, too, is (~statistically~)
correlated with the sites of SNR from reasons of energetics.

Of importance is the fact that the positions of the peak intensities
in the maps at 5 GHz, 10 GHz, 60$\mu$ (~IRAS~), 10 GHZ, 5 GHZ and 2.7
GHz (~Effelsberg: http://www.mpifr.\\de/old\_mpifr/survey.html~) are
all coincident with the 'peaks' in the CR plot (~Figure 2~). They are
not coincident with the peaks in the column density of gas (~Figure 1~).

The likelihood of the radio flux being an indicator of the distribution of SNR (~'present' and
past~) in the GCR is enhanced by the linear size
distribution of the excesses in Figure 2, as will be demonstrated
later. 

\section{Analysis of the data}

\subsection{General Remarks}

We start with Figure 1. It is interesting, and perhaps very
significant, to note the large discrepancy in the region between $l =
+1^\circ$ and $l = +2^\circ$ between the column density of gas and the
measured gamma ray intensity - a discrepancy that would not exist if
the CR intensity were constant. In other words, there is a significant 
fraction of the gas in the GCR (~some 18\%~) that is under-populated
by CR. It is this mass of gas, beyond $l = 1^\circ$, that makes the
gas non-symmetric about the Galactic Centre. The question of symmetry
in the various GCR properties is taken up again later.

Moving to Figure 2, a number of remarks can be made.
\begin{itemize}
\item[(i)] The likelihood of systematic errors in the inferred
column density of molecular hydrogen means that the absolute
values of the enhancement in CR intensity ratio  (~denoted $R(l)$~) 
are uncertain. However, the shape of the longitude-dependence 
should be reasonable.

\item[(ii)] There should be a trend of diminishing R ($l$) with
increasing $|l|$ if our contention about diffusion of CR from
sources in SgrA* is correct.

\item[(iii)] Some measure of correlation of R ($l$) with the
integrated radio intensity should occur if, as has been
remarked, there are SNR in the GCR itself. An alternative way
of approaching this problem is in terms of CR diffusion - in high
B regions there will be a smaller diffusion coefficient because of
increased turbulence.
\end{itemize}

\subsection{Overall CR excess}

Inspection of Figure 1 shows that the gamma ray excess falls down
slightly when receding from the Galactic Centre, ie with increasing
$|l|$ and, correspondingly, the CR intensity also falls. 
Certainly the GCR is a singular region with unique properties. Any
extrapolation of its characteristics to the wider longitude range
should be taken with care. However, taking into account the large errors
in the observed intensity of the diffuse gamma rays and uncertainty in
the column density of the gas we cannot rule out a weaker longitude
dependence of the CR intensity in the wider longitude range of the
Inner Galaxy, than is indicated in Figure 2.

The analysis of the diffuse gamma-ray profile at lower GeV energies
obtained with the SASII satellite indicated a weak radial CR gradient 
in the Inner Galaxy, (~Issa and Wolfendale,1981~), as did the CGRO
satellite (~Strong and Mattox,1996~). The CR intensity in
the Inner Galaxy does not exceed the local value by a factor of more
than 1.2. If our latter value
of the excess equal to 1.83$\pm$0.15 can be extrapolated into a wider
longitude region than $|l| < 1.6^\circ$, it might mean that the CR
gradient for sub-TeV and TeV energies is stronger than for GeV
energies, as has been predicted by us (~Erlykin and Wolfendale, 2002~). 

\subsection{The GC Region}

Our model, to be tested, comprises SN which have exploded in the
central region (~ together, probably, with the 'normal quota' of SNR associated
with the gas in general~), providing protons which diffuse away and permeate
the molecular gas, in which they produce secondary pions and thereby 
gamma rays. It is supposed that recent SN provide the peak at the
position of SgrA* itself (to be discussed later).

In Figure 4 we show the gamma-ray profile expected from a single SN
which exploded either $10^4$ or $10^5$ years ago. Calculations have
been made for both `normal' and `anomalous' diffusion (~the distinction relates to the 
nature of the 'scattering centres' in the interstellar medium~). Briefly, the 
difference between them can be reduced to two basic features: \\
(i) the diffusion radius $R_d(t)$ for anomalous diffusion depends linearly on time as
$R_d(t) \propto t$ while for normal diffusion $R_d(t) \propto \sqrt{t}$. Both radii
are normalized to 1 kpc at the time equal to the mean life time of CR particles for the
particular energy; \\
(ii) the lateral distribution function of the CR density $\rho(r)$ for anomalous 
diffusion is not gaussian as for normal diffusion, but has a more complicated shape: 
 $\rho(r) \propto (1+(\frac{r}{R_d})^2)^{-2}$. \\ 
The difference between these two diffusion modes is discussed in more detail in 
Lagutin,2001, Erlykin and Wolfendale, 2002, Erlykin, Lagutin and Wolfendale, 2003.

The effective averaged gas density in our calculations for the region in question 
$|l| < 1.6^\circ, |b| < 0.3^\circ$ is $100 cm^{-3}$ (~this
being much less than the high values of $\sim 10^4 cm^{-3}$ quoted because of
the presence in our studied volume of a large volume of low density
material and the 'filling factor', ie the fraction of space occupied by the high 
density molecular gas~). 
\begin{figure}[htbp]
\begin{center}
\includegraphics[height=15cm,width=8cm,angle=-90]{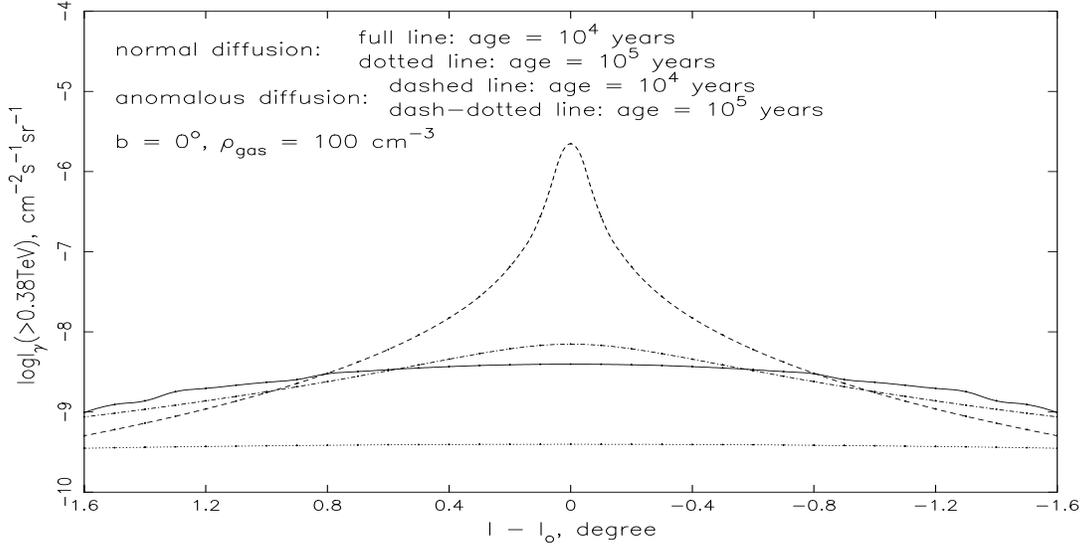}
\caption{\footnotesize Longitude dependence of the gamma ray intensity
from a single source at longitude $l$, in a medium of density $100
cm^{-3}$ at $b = 0^\circ$. Two ages are considered, $10^4$ and $10^5$
years and two modes of diffusion: normal and anomalous.}
\label{fig:fig4}
\end{center}
\end{figure}

It is seen that there is no sharp
peak in the centre even for the youngest SN with an age of $10^4$
years. The fall of the intensity with longitude in the wings is too strong
to be compatible with the experiment. If we reduce the age below
$10^4$ years the sharpness of the peak increases but the fall of the
intensity in the wings increases too. Therefore, to get agreement
with experiment both in the central peak and in the wings ( Figures 1 and 
2 ) we cannot use just a single SN and need the succession of SN
explosions distributed in time.

In Figure 5 we show the results of the calculation for a succession
of SN in SgrA*. Two sets of SN rates are taken:
$10^2$ SN in $10^4$ years and $10^3$ SN in $10^5$ years with the uniform temporal 
distribution of SN explosions within these time intervals. Since the average rate in 
both cases is 1 SN in 100 years the diffusion approximation for CR from some young SN 
and large distances from the GC cannot be valid because the CR velocity cannot exceed 
the speed of light. We have introduced limitations for this effect and have found that the 
gamma-ray intensity at $\ell = 2^\circ$ decreased by $\sim$1.6\%. In all subsequent 
calculations we applied these limitaions.
\begin{figure}[htb]
\begin{center}
\includegraphics[height=15cm,width=8cm,angle=-90]{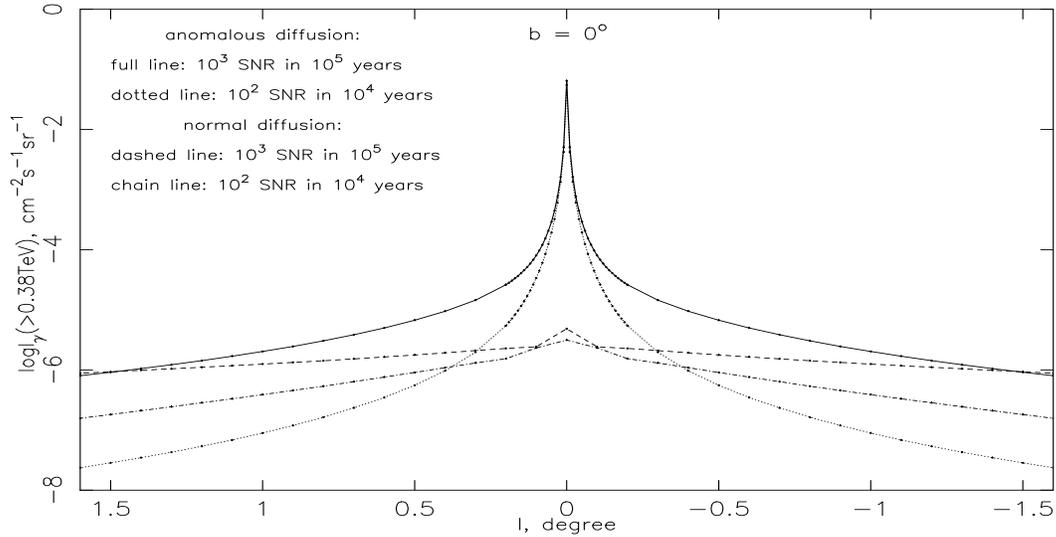}
\caption{\footnotesize The angular profile of the gamma ray intensity
in the GCR predicted for our model with $n = 100 cm^{-3}$ and either
$10^3$ SN in $10^5$ years at the position of SgrA*, placed at the
longitude $\ell = 0^\circ$, or $10^2$ SN in $10^4$ years. Results are
given for boh anomalous and normal diffusion.}
\label{fig:fig5}
\end{center}
\end{figure}

In Figure 5 the (~important~) sharp spike, which results
mainly from SNR younger than about 10$^4$ years, from which the CR do
not diffuse very far, is a consequence of anomalous diffusion. Normal 
diffusion gives a much weaker spike (~see also Figure 4~).
In practice, in view of the sources being in SgrA* itself, where the
density is very much higher than 100$cm^{-3}$, the peak will be even higher.
The other, later SN will have given particles which have diffused out
to permeate more of the molecular material in the GCR.

The extent to which the idealised 'lateral 
distributions' in Figure 5 would be modified using the actual column
density of gas (~rather than gas of constant density~), averaging over the galactic 
latitude interval $|b| \leq 0.3^\circ$ and applying the correction for the finite 
angular resolution of the telescope of $0.1^\circ$, can be seen in
Figure 6. The general trend is seen to be preserved. For the comparison we also provide
 the results of HESS observations. The difference between the absolute values of 
calculated and observed intensities is discussed below.  
\begin{figure}[htbp]
\begin{center}
\includegraphics[height=15cm,width=8cm,angle=-90]{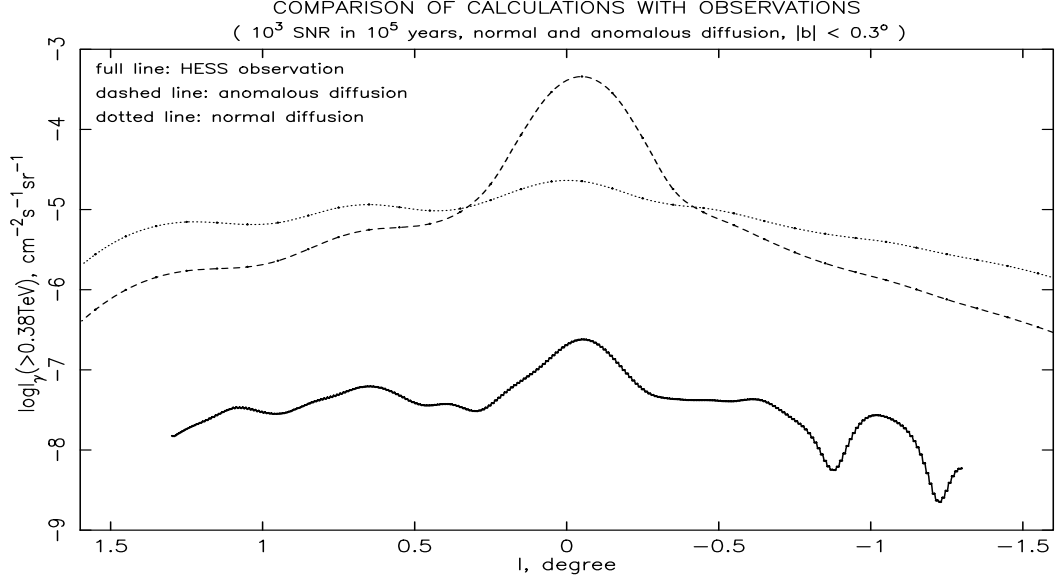}
\caption{\footnotesize Angular profile of gamma ray intensity for the
whole GCR for $10^3$ SNR in $10^5$ years and for the actual molecular gas distribution 
of Figure 1 (~allowed for in an approximate way~). The results of calculations are
averaged over the galactic latitudes $|b| \leq 0.3^\circ$ and smoothed taking into 
account the finite angular resolution of the telescope of $0.1^\circ$. Normal and 
anomalous diffusion are presented by dotted and dashed lines respectively. The 
results of HESS observations are also shown by the full line for comparison.} 
\label{fig:fig6}
\end{center}
\end{figure}

\section{Interpretation of the results}

\subsection{General Remarks}

Guidance as to the frequency of SN comes from remarks in a variety
of works (~eg I~) that of order $10^2$ SN in $10^4$ years to $10^3$ SN over
the past $10^5$ years would have been sufficient to provide the
energy necessary for the very strong wind and other features
visible in this unique region of the Galaxy. Production over a longer
period seems unlikely, although it must be said that there is some
evidence for a bout of star formation between 3$\cdot 10^6$ and 7$\cdot 10^6$
years age (~Krabbe et al., 1995~).

It must be admitted that the manner in which the particles diffuse in this region is
debatable both by way of the diffusion coefficient to adopt and
the manner of diffusion, viz `normal' or `anomalous'. We consider
that, in view of the disturbed conditions in the region and highly
non-uniform distribution of gas, the
mode of diffusion in the very central region at least will be
'anomalous'  (Erlykin, Lagutin and
Wolfendale, 2003). Concerning the diffusion coefficient, in the
absence of clear information we adopt the `local' value
(pertaining to the Galaxy as a whole).

\subsection{SgrA* alone}

It is easiest to consider this region alone to start with. With
only one SN, young enough that the particles are sufficiently
confined to the region as to give gamma rays which are well within
the point spread function (~PSF~) of the detector $\simeq 0.1^\circ$,
the predicted flux can be taken from our earlier work (Erlykin,
Wolfendale, 2003)
\begin{equation}
F_{\gamma} (> 1~TeV) = 35\cdot 10^{-12} \left(\frac{E_{o}}{10^{51}
erg}\right)\left(\frac{n}{1~cm^{-3}}\right)\left(\frac{d}{1
kpc}\right)^{-2} \Delta \cdot f~~ cm^{-2}s^{-1}
\end{equation}
Here, for a start, we adopt for the SN explosion energy $E_{o} = 10^{51}~erg$, the
gas density $n = 100~cm^{-3}$, the distance to the GC $d = 8.5~kpc$ and the fraction 
of the SN explosion energy transferred to CR,
$\Delta  = 0.1$, as usual (Berezhko et al., 1996).  $f$ is the
efficiency-factor, which is actually a ratio of the observed gamma-ray intensity to 
that expected for a standard model of the CR production, to be determined from the 
observations. $f$ can, of course, be much less than unity if the physical conditions in
 the SNR differ considerably from the conventional ones, resulting in weaker shocks, 
etc. The result is $F_{\gamma} (> 1 TeV) = 5f \times 10^{-12}~cm^{-2}~s^{-1}$.

Following the HESS work, we adopt 0.38 TeV as the threshold energy
and, using the measured spectrum of gamma rays (with differential
exponent $\gamma = 2.3$), the observed total flux from the GC
including the point source at SgrA* is \\
$ F_{\gamma} (> 0.38 TeV) = 24 \times 10^{-12}~cm^{-2}~s^{-1}$ \\ 
$ F_{\gamma} (> 1 TeV) = 7\times 10^{-12}~cm^{-2}~s^{-1}$ \\
Taken at face value, this would require, for one SN alone, $f \simeq 1.4$.
Thus, for 100 or 1000 SN we would need $f \approx 10^{-2}$ or $10^{-3}$
It is evident that there is no shortage of energy in the SNR
hypothesis.

Later observations have given slightly different values of the fluxes and threshold 
energy but our arguments are unchanged.

Figures 5 and 6 allow us to make a more accurate estimate. The flux from the
spike within $|l| < 0.2^\circ$ would qualify as that from the discrete
source insofar as its finite width would not be detected in the
presence of a PSF with the half width at half maximum of $0.1^\circ$ 
(~which has a wide tail~).

Of the two situations considered, it is evident that 10$^2$ SN in
10$^4$ years gives too steep a 'lateral distribution' (~comparing
Figure 5 with Figure 2~) but the shape for 10$^3$ SN in 10$^5$ years
is close. Here, for $|l| < 0.2^\circ$, we predict a flux of $5\cdot
10^{-9} cm^{-2}s^{-1}$ for the case of anomalous diffusion. The observed flux is 
$7\cdot 10^{-12} cm^{-2}s^{-1}$ so that $f \simeq 1.4\cdot 10^{-3}$ for this
case.

\subsection{The Galactic Ridge}

As remarked already 
it is evident that the fall of predicted
gamma ray intensity with increasing longitude is about right to
explain the gamma ray profile for $10^3$ SN in $10^5$ years. 
There is a problem, however, in that
the 'inner wings' of the predicted distribution, from $|l|: 0.05^\circ$ to 
$0.4^\circ$, are too strong. A likely explanation here is in terms of a
non-uniform temporal distribution. An estimate of the expected flux in the Ridge can
be derived from Figure 6: it is $8.6\cdot 10^{-9} cm^{-2}s^{-1}$. With
the efficiency factor of $1.4\cdot 10^{-3}$ derived for the sources in
SgrA* we have $12\cdot 10^{-12} cm^{-3}s^{-1}$, to be compared with the
observed $17\cdot 10^{-12} cm^{-2}s^{-1}$. The result is 
close, although we have to admit that many parameters involved in the calculations 
are uncertain.

This is where the SNR, which undoubtedly must be present in the Ridge
material itself, may assume importance.  Their likely
number can be estimated. as follows. For the Galaxy as a whole, with 
molecular mass $M(H_{2})
\sim 10^{9}~M_{\odot}$ (~Dame et al., 2001~) and SN rate
$10^{-2}~y^{-1}$, we have $\sim
10^{-11}~M_{\odot}^{-1}~y^{-1}$. It is likely that this value is
also appropriate to the Ridge region so that in
$10^{5}~y$, with a mass of $4.4 \cdot 10^{7}~M_{\odot}$,  we expect
44. In fact, the number may be nearer 10 in view of the dependence of
SNR density on column density of molecular hydrogen being slower than
linear, more nearly to the power 0.6 (~using the summary of Fatemi and
Wolfendale, 1996 for SNR, pulsars and $N(H_2)$~). Assuming that these 
SN give the same contribution to the flux in
the wings as the central SN and using the calculated flux in the Ridge
for 1000 central SN of $8.6\cdot 10^{-9} cm^{-2}s^{-1}$ and observed
flux of $1.7\cdot 10^{-11} cm^{-2}s^{-1}$, the derived $f$
- value is $\sim 0.1$. 

A measure of validity for the view of a significant, or even major,
contribution from the 'conventional' SNR comes from the CR-radio
correlation (~Figure 3 and \S3.3~). Three of the diffuse radio
sources appear in the SNR catalogue of Green (~2000~) and in general
we would expect the radio map to be a good indicator of past and
present SNR. A similar situation arises for the radio map and
identified SNR for the Large and Small Magellanic Clouds (~Mills and
Turtle,1984~). A natural explanation would exist for the 'bumps' in
Figure 2 in terms of SNR, in view of the expected profiles of
CR-produced gamma rays shown in Figure 4; SNR of (~typical~) age a few 
10$^4$ years would have profiles
similar to the 'bumps' in Figure 2. Specifically, the mean half-width
of the bumps which can be resolved is $\sim 0.24^\circ$ and this is
just what would be expected for about 10 SN having exploded in the
last $10^5$ years.

So far, therefore, there seem to be two possibilities to explain the
results. Firstly, there were $10^3$ SN in $10^5$ years in SgrA*, the
CR diffusing through the Ridge causing the Ridge emission and the
recent SN giving CR very close to SgrA* which caused the point
source. The problem is that the needed gas density to get the observed
ratio of Ridge flux to SgrA* 'point source' flux is very low, by SgrA*
standards.

Secondly, perhaps the SNR in the Ridge itself were responsible for
the CR there. The number of SN in SgrA* could then be smaller, with
higher density gas allowed. The problem with the high intensity predicted
in the range $l: 0.05^\circ$ to $3^\circ$ or $4^\circ$ would then be minimized - 
it would be due simply to a statistical fluctuation.

\subsection{The efficiency-factor, $f$}

There are many phenomena  which could contribute to $f$ being less then
unity for the unusual conditions in the GCR. These are, mainly,
\begin{itemize}
\item[1.] The possibility that the majority of SN there are not of
Type II - the main sources of SNR which accelerate CR to very high
energies. 
\item[2.] The high gas density causes the Sedov radius
(which is proportional to $n^{-\frac{2}{5}}$, Axford, 1981) to be
small; specifically it falls to only a few pc. The time taken to
reach this radius, after which CR acceleration is reduced, is
probably too short for efficient acceleration, despite the
increased magnetic field in the GCR.
\item[3.] The high gas density probably causes the injection
efficiency to be low (~Drury et al., 1996 and private communication~), the point being that
ionization losses will be considerable during injection for the
sub-relativistic particles.
\item[4.] The tube-like magnetic fields, referred to in \S 1, which
will convey particles out of the Galaxy.
\item[5.] A very likely effect relates to the Galactic Wind. This
is currently very strong in the GCR (eg Breitschwerdt et al.,2002;
V\"{o}lk and Zirakashvili,2004; Sunyaev et al 1993) - with a velocity
of a few thousand $km~s^{-1}$.
\end{itemize}

Concerning item 1, it seems unlikely
that there is a shortage of the necessary massive pre-SN
stars. Indeed, in parts
(eg SgrB2), there are unusually massive stars being produced
'furiously' (~I~). Presumably the high rate of
SN production overall in the GCR, and particularly in SgrA*, gives
rise to many Type II SN.

The other factors are, therefore, considered to be the
relevant ones.

\subsection{The Cosmic Ray Gradient}

A well known feature of the CR distribution in the Galaxy as a whole - as
inferred from gamma ray data - is the smallness of the `CR
gradient', i.e. the dependence of CR intensity on Galactocentric
radius, R (eg Erlykin et al., 1996; Hunter, 2001.) and we believe that
this has relevance, here. As is well known, 
the gradient for CR is far less than that for the (assumed) parent SNR.
  A number of possibilities have been put forward,
involving, for example, Galactic winds (~Breitschwerdt et al.,2002;
V\"{o}lk and Zirakashvili,2004~) and radial dependent diffusion
coefficients (~Erlykin and Wolfendale, 2002~).

The results from the present analysis of the GCR lead to our
suggesting that the efficiency of CR production by SNR compared with the standard model
where 10\% of the explosion energy is transformed into CR, is dependent on the nominal 
gas density of the ambient medium into which the SN expands. It will
be apparent from the start that the low implied CR intensity at
the GC despite the large number of SNR is in the spirit of a gas
density dependent SNR efficiency.

In Figure 7 we present the `CR efficiency', given here as the
ratio of inferred CR intensity, I(CR) to the surface density of
SNR (as used by us previously, Erlykin and Wolfendale, 2005)
versus the surface density of molecular hydrogen, $\sigma
(H_{2})$, as well as the GC values. The fall-off of efficiency with 
$\sigma(H_2)$ is very marked.

\subsection{The preferred model}

At this stage it is not possible to decide between the two models but
the mixed model (~ie SN in both SgrA* and the Ridge~) appears more
likely. The $f$-values for the two situations are given in Figure 7. 
Consideration can be given to each region in turn.

{\bf SgrA*}. One hundred SN is probably the smallest that can be
allowed, and able to give the necessary energy injection for the high
wind, plasma temperature, etc. Thus, for one thousand
SN the figure is $1.4\cdot 10^{-3}$ ( not much smaller because much
of the CR energy escapes from the central region~).

{\bf The Galactic Ridge}. For 10 sources, we have $f\simeq 0.1$.
The values are plotted in Figure 7. The abscissae are illustrative in
that they are the approximate surface densities of gas for the
regimes in question.
Also shown in Figure 7 is the result for the source G0.9+0.1, which
appears to be due to a single SNR. It should be remarked that the
region in which the SN exploded is outside the region of high gas
density and thus the efficiency might be expected to be comparatively
high.
\begin{figure}[htbp]
\begin{center}
\includegraphics[height=15cm,width=8cm,angle=-90]{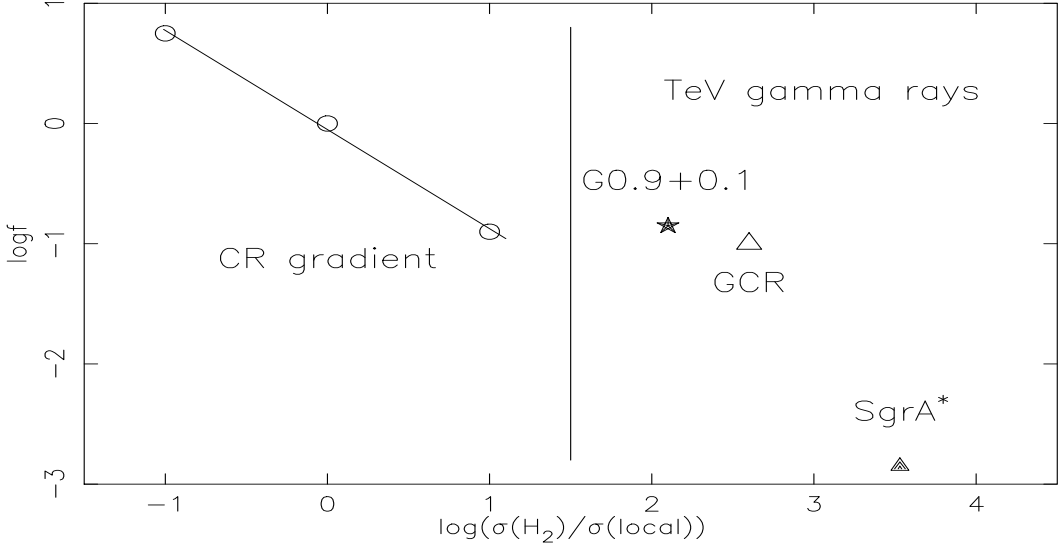}
\caption{\footnotesize Efficiency of SNR for accelerating CR of
energy and intensity sufficient to give gamma rays of GeV energy (~CR
gradient results~) and above 0.38 TeV as a
function of the surface density of molecular hydrogen. The local
values of efficiency and surface density of gas are taken as datum in
each case. For the TeV gamma rays we adopt a nominal mean density of
100 $cm^{-3}$.}
\label{fig:fig7}
\end{center}
\end{figure}

\subsection{CGRO data in the GeV region}

It is relevant to see to what extent the results reported here have
correspondence in the GeV region. To this end we have examined the CGRO
observations reported by Hunter et al. (~1997~); these results referring to
energy ranges from 30/100 MeV to $E_\gamma > 1000$ MeV. Considering
the results for $|b| < 2^\circ$, the region for $|l| < 1^\circ$ has a
spectrum flatter than the surroundings by $\Delta \gamma = 0.15$,
when attention is devoted to the contribution from interactions with
molecular hydrogen alone. This is in the spirit of that reported here
for HESS, where the spectrum of gamma rays in the GCR is flatter than average, by
$\Delta \gamma \simeq 0.31 \pm 0.22$. Bearing in mind the inferior
angular resolution of CGRO and the quoted 'errors', the difference in
the $\Delta \gamma$-values is understandable.

Of interest, too, is the fact that there are small gamma-ray excesses for $|b|
< 2^\circ$, $E_\gamma > 1000$ MeV at $l$-values where there are also
peaks in the 408 MHz maps of Haslam et al. (~1981~). Specifically, of
23 peaks in the radio map above a consistent height (~50K compared
with a central region of height $\sim$450K and an Anti-Centre level of
 $\sim$50K~) some 14 coincide with gamma ray excesses. Thus, the GCR
correlation of gamma-ray excesses with radio intensity seems to have a
counterpart at GeV energies. 

\section{Discussion}

The HESS group appear not to have analysed the results from as detailed a
standpoint as we ourselves; instead, they take an empirical approach
by way of fitting a Gaussian to the cosmic ray distribution (~see
Figure 1~) in GCR and a point source in the GC itself. This can
perhaps be regarded as a zero order approach.

Some general remarks about the analysis are in order. In \S1 we
discussed the general energetics of the GCR and pointed out the very
high - and nearly equal - energy densities of the major
components. The cosmic ray energy density is clearly lower by 3
orders of magnitude. Of the discrepancies, that between the magnetic
field and the CR energy density is, at first sight, hardest to
understand. Locally (~ie in the solar system~) there is equality and
good reasons have been put forward in terms of trapping, but this
must break down in the GCR. Presumably, the reason is the field
geometry, ie the outward- directed nature of the high field
flux-tubes, which helps to lose CR, unlike locally, where the magnetic
field acts to contain them.

Another topic needing discussion is that relating to symmetry - and
lack of it. Inspection of Figure 1 shows the distinct lack of symmetry
of the density of molecular gas about an axis through the Galactic
Centre along $l = 0^\circ$. Specifically, the symmetry in the column
density versus longitude plane (~Figure 1~) is about $l = +0.5^\circ$.
This is in sharp contrast to that for the CR intensity, (~Figure 2~),
which is about $l = 0.0^\circ$ and, indeed, the gamma ray intensity
itself (~Figure 1~), which is about $l = +0.1^\circ$. The radio
intensity (~Figure 2~) is also symmetrical about $l = 0.0^\circ$. Of
particular interest is the map for 60$\mu$ (~Uchida et al.,1996~),
this radiation arising from hot dust; for this map there is symmetry
about $l = 0.0^\circ$. Most importantly, the 60$\mu$ has almost
disappeared by $l = 1.0^\circ$, clearly, the dust in the anomalous
region, $l: 1^\circ$ to $2^\circ$, is 'cold'. This result is in accord
with our contention that the CR intensity is very low there, by virtue
of the (~chance~) paucity of recent SN.

\section{Conclusions}

The main results can be summarised thus. \\
1. Supernova remnants can be invoked to explain the HESS results for
the Galactic Centre Region. The rate for the SgrA* region, $10^3$ SN in
$10^5$ years, as recommended by other
work (~I~), gives a consistent picture, although somewhat fewer, taken
with 'conventional' SNR in the Ridge, is preferred.\\
2. The 'consistent picture' requires particularly low SNR
efficiency-factors for SgrA* (~but less so for the Ridge if
'conventional' SNR play a role~). Taken together
with results from the well-known small 'CR-gradient' in the Galaxy as
a whole, they lead to a satisfying smooth dependence of efficiency factor
on local molecular density. The ideas put forward in \S5.4 would be
expected to give such a dependence. \\
3. An interesting feature is the correlation of CR intensity with
smoothed radio intensity. We regard this as providing support for the
idea that SN in the Ridge are important. \\
4. The CGRO data in the sub-GeV and GeV region are consistent insofar
as they, too, give a somewhat flatter gamma ray spectral exponent for
the GCR than for the rest and there are correlations of intensity
excesses with spikes in the radio map. \\
5. The very high magnetic field inhibits high energy electrons,
thereby enabling conclusions about protons (~more accurately, nuclei~) 
to be drawn.

{\bf Acknowledgments}

One of authors (ADE) thanks the Royal Society and the University of Durham
for financial support. We also thank L.O'C.Drury, T.Bell, P.M.Chadwick and J.L.Osborne 
for helpful comments. Remarks by the referees of a previous version of this paper were 
particularly useful; we thank them.

{\bf References}

\noindent Aharonian, F.A. et al., Astron. Astrophys., {\bf 425}, L13 (2004) \\ 
Aharonian, F.A. et al., Nature, {\bf 439}, 695 (2006) \\
Albert J. et al., Astrophys. J., {\bf 638}, L101 (2006) \\
Altenhoff, W.J. et al., Astron. Astrophys. Suppl. Ser., {\bf 35}, 23 (1979) \\
Axford, W.I., Proc. 17th Int. Cosmic Ray Conf. (Paris), {\bf 12}, 155
(1981) \\
Bania, T.M., Astrophys. J., {\bf 216}, 381 (1977) \\
Berezhko E.G. et al. JETP, {\bf 88}, 1 (1996) \\
Biermann P.L., Proc. 23d Int.Cosmic Ray Conf., Calgary, Inv. Rapp. and
High. Papers, 45 (1993) \\
Breitschwerdt D. et al., Astron. Astrophys., {\bf 385}, 216 (2002) \\
Dame, T.M., Hartmann Dap and Thaddeus, P., Astrophys. J., {\bf 547},
792 (2001) \\ 
Drury, L.O'C. et al., Astron. Astrophys., {\bf 309}, 1002 (1996) \\
Erlykin, A.D. and Wolfendale, A.W., J. Phys. G: Nucl. Part. Phys.,
{\bf 27}, 941 (2001); J. Phys. G: Nucl. Part. Phys. {\bf 28}, 2329
(2002); J.Phys.G: Nucl. Part. Phys., {\bf 29}, 641 (2003); 
Proc. 29th ICRC, Pune, {\bf 3}, 133 (2005) \\
Erlykin, A.D. et al., Astron. Astrophys. Suppl. Ser., {\bf 120}, 397
(1996) \\
Erlykin, A.D., Lagutin, A.A. and Wolfendale, A.W., Astropart. Phys.,
{\bf 19}, 351 (2003) \\
Fatemi, S.J. and Wolfendale, A.W., J. Phys. G: Nucl. Part. Phys., {\bf
 22}, 1089 (1986) \\ 
Green D.A., 'A catalogue of galactic supernova remnants (~2000 August
version~)', University of Cambridge (2000) \\
Handa,T. et al., PASJ, {\bf 39}, 709 (1987) \\
Hartman R.C. et al., Astrophys. J. Suppl. Ser.,{\bf 123}, 79 (1999) \\
Haslam, C.G.T., Salter, C.J. and Stoffel H., 'Origin of Cosmic Rays',
ed. by G.Setti, G.Spada and A.W.Wolfendale, IAU, 217 (1981) \\
Hunter, S.D. et al., Astrophys. J., {\bf 481}, 205 (1997) \\
Hunter, S.D., 'High Energy Gamma Ray Astronomy' ed. by F.A.Aharonian
and H.J.V\"{o}lk, American Institute of Physics, I-56396-990-4, p.171
(2001) \\
Issa, M.R. and Wolfendale A.W., J. Phys. G: Nucl. Part. Phys., {\bf
7}, L187 (1981) \\
Kosack K. et al., Astrophys. J., {\bf 608}, L97 (2004) \\
Krabbe, A. et al., Astrophys. J. Lett., {\bf 447}, L95 (1995) \\
Lagutin A.A. et al., Nucl. Phys. B (~Proc.Suppl~), {\bf 97}, 267 (2001);
Proc. 27th ICRC, Hamburg, {\bf 5}, 1900 (2001) \\                     
Mayer-Hasselwandler H.A. et al., Astron. Astrophys., {\bf 103}, 164
(1982); Astron. Astrophys. {\bf 335}, 161 (1998) \\
Mills, B.Y. and Turtle, A.J., 'Structure of the Magellanic Clouds',
eds. S.Vander Berrgh and K.S. de Beer, IAU Suymp., {\bf 108} (1984) \\
Morris, M. and Serabyn, E., Ann. Rev. Astron. Astrophys., {\bf 34},
645 (1996) \\
Oka, T. et al., astro-ph/9810434 (1998) \\
Strong A.W. and Mattox J.R., Astron. Astrophys., {\bf 308}, L21 (1996) \\
Sunyaev, R., Markevitch, M. and Pavlinsky, M., Astrophys. J., {\bf
407}, 606 (1993) \\ 
Sveshnikova L.G. Astron. Astrophys., {\bf 409}, 799 (2003) \\
Thompson, D.J. et al., 'The Structure and Content of the Galaxy and
Galactic Gamma Rays', p.1, GSFC, X-662-76-154 (1976) \\
Tsuboi, M., Handa, T. and Ukita, N., Astrophys. J., {\bf 120}, 1
(1999) \\
Tsuchiya K. et al., Astrophys. J., {\bf 606}, L115 (2004) \\
Uchida K.I. et al., Astrophys. J., {\bf 462}, 768 (1996) \\
V\"{o}lk H.J. and Zirakashvili V.N., Astron. Astrophys., {\bf 417},
807 (2004) \\
Wolfendale, A.W., Quart. Journ. Roy. Astr. Soc., {\bf 24}, 122 (1983)
\end{document}